%% file: PgNN.tex
\documentclass[10pt,journal,compsoc]{IEEEtran} % 源代码
\usepackage{amsmath}
\usepackage[pdftex]{graphicx}%only allow the pic placed at the top
\usepackage{dblfloatfix}% allow the pic placed at the bottom %https://tex.stackexchange.com/questions/167186/figure-environment-skips-page-while-using-two-column-document
\DeclareGraphicsExtensions{.pdf}
\usepackage{epstopdf}

\ifCLASSOPTIONcompsoc
  \usepackage[nocompress]{cite}
\else
  \usepackage{cite}
\fi

\begin{document}

%% The artical title part
\title{PgNN: Physics-guided Neural Network for Fourier Ptychographic Microscopy}

%% The author part
\author{\textbf{~Yongbing~Zhang,~Yangzhe~Liu,~Xiu~Li,~Shaowei~Jiang,~Krishna~Dixit,~Xinfeng~Zhang, ~and~Xiangyang~Ji}% <-this % stops a space
\IEEEcompsocitemizethanks{\IEEEcompsocthanksitem Y. Zhang, Y. Liu and X. Li are with the Graduate School at ShenZhen, Tsinghua University, ShenZhen 518055, China. %\protect\\E-mail: zhang.yongbing@sz.tsinghua.edu.cn.
\IEEEcompsocthanksitem S. Jiang and K. Dixit are with the Department of Biomedical Engineering, University of Connecticut, Storrs, CT 06269, USA.}}
%\IEEEcompsocthanksitem X. Zhang is with the School of Computer Science and Technology, University of the Chinese Academy of Sciences, Beijing 100049, China. \protect\\E-mail: zhangxinf07@gmail.com.
%\IEEEcompsocthanksitem X. Ji is with TNLIST and Department of Automation, Tsinghua University, Beijing 100084, China. E-mail: xyji@tsinghua.edu.cn.}% <-this % stops an unwanted space

%% The paper headers
\markboth{}%
{Shell \MakeLowercase{\textit{et al.}}: Bare Demo of IEEEtran.cls for Computer Society Journals}

%% 0.abstract
\input{0.abstract.tex}
%% 1.Introduction
\input{1.introduction.tex}

%% 2.related work
\input{2.relatedwork.tex}

%% 3.background
\input{3.background.tex}
%% 4.method
\input{4.method.tex}

%% 5.experiment

\input{5.experiment.tex}

%% 6.7.discussion and summary

\input{6.7.discussion_and_summary.tex}
%% 8.reference
%\input{sections/8.reference.tex}

%% addition: The figure part
%% Fig.1 is declared in beginning of section 3

%% Fig.2;3 is declared in the beginning of the section4/Ajustable Performance

%% Fig.4 is is declared in the beginning of the section4/Ajustable Performance; one page before we want

%% Fig.5 is is declared in the beginning of the section4/Improved Modality of Aberration; one page before we want

%% Fig.6 is is declared in the beginning of the section5

%\ifCLASSOPTIONcompsoc
%\section*{Acknowledgments}
%\else
%\section*{Acknowledgment}
%\fi


\bibliographystyle{IEEEtran}      %LaTex Class文件, IEEEtran为给定模板格式定义文件名
\bibliography{IEEEabrv,Reference}   
\end{document}

%% file: 0.abstract.tex
%%--------------------------------------
%% The abstract and index terms
\IEEEtitleabstractindextext{%
\begin{abstract}
Fourier ptychography (FP) is a newly developed computational imaging approach that achieves both high resolution and wide field of view by stitching a series of low-resolution images captured under angle-varied illumination. So far, many supervised data-driven models have been applied to solve inverse imaging problems. These models need massive amounts of data to train, and are limited by the dataset characteristics. In FP problems, generic datasets are always scarce, and the optical aberration varies greatly under different acquisition conditions. To address these dilemmas, we model the forward physical imaging process as an interpretable physics-guided neural network (PgNN), where the reconstructed image in the complex domain is considered as the learnable parameters of the neural network. Since the optimal parameters of the PgNN can be derived by minimizing the difference between the model-generated images and real captured angle-varied images corresponding to the same scene, the proposed PgNN can get rid of the problem of massive training data as in traditional supervised methods. Applying the alternate updating mechanism and the total variation regularization, PgNN can flexibly reconstruct images with improved performance. In addition, the Zernike mode is incorporated to compensate for optical aberrations to enhance the robustness of FP reconstructions. As a demonstration, we show our method can reconstruct images with smooth performance and detailed information in both simulated and experimental datasets. In particular, when validated in an extension of a high-defocus, high-exposure tissue section dataset, PgNN outperforms traditional FP methods with fewer artifacts and distinguishable structures.
\end{abstract}

\begin{IEEEkeywords}
deep learning, Fourier ptychography, super-resolution imaging, physics-guided neural network
\end{IEEEkeywords}}

\maketitle
\IEEEdisplaynontitleabstractindextext
\IEEEpeerreviewmaketitle
%%--------------------------------------

%% file: 1.introduction.tex
\IEEEraisesectionheading{\section{Introduction}\label{sec:introduction}}
\IEEEPARstart{I}{n} many biomedical applications, it has always been difficult to simultaneously obtain both high resolution and large field of view (FOV) images. Regardless of advancements in sophisticated mechanical scanning microscope systems and lensless microscopy setups, the modification of conventional microscopes to obtain ideal high-resolution results has been a hot topic of recent research work. Fourier ptychography (FP) in particular is a simple and cost-effective analytical method for this application \cite{1},~\cite{2},~\cite{3},~\cite{New_1},~\cite{New_2}.%

As a newly proposed computational imaging method, FP integrates principles of phase retrieval \cite{4,New_5.1,New_5.2,34} and aperture synthesizing \cite{6,7,8,9} for achieving both high resolution and large FOV images with amplitude and phase. By introducing a programmable color LED matrix as an angle-varied coherent illumination source, FP can sequentially capture low-resolution intensity images corresponding to distinct apertures of the frequency spectrum of the sample. This capturing mechanism is able to shift the high-frequency spectrum into the passband of the low numerical aperture (NA) objective lens and makes it possible for low-NA lens to detect such information. Inspired by aperture synthesizing, FP then iteratively stitches a series of captured low-resolution intensity images together in the Fourier domain to enlarge the passband of the microscopy while no interferometric measurements for phase detection are needed. To further recover the lost phase component, FP applies the phase retrieval mechanism to unroll relevant information from sequential intensity images. As such, FP can reconstruct a object with an equivalent high-NA and a larger FOV via a conventional microscope and is widely used in hematology \cite{10}, pathology \cite{11} and so on. 

FP is such a low-cost and effective approach that it has received increasing attention over the past few years. For the image reconstruction procedure, Ou \emph{et al.} \cite{2} embedded a pupil function recovery process in FP to reconstruct the optical aberrations produced by the unknown point spread function (PSF) of the objective lens, resulting in greater flexibility in FP. For the hardware setup, the strategic application of multiplexed illumination effectively reduced the acquisition time and image capturing requirements~\cite{3},~\cite{12},~\cite{13}. 

Recently, deep neural networks (DNNs) have been proven to reliably provide inductive answers to the inverse problem in computational imaging \cite{8}. Thus, the ability of FP to reconstruct the sample image can be enhanced by seeking cross-integration with deep learning. In early studies, several deep learning architectures, such as CNN and cGAN, have been applied to solve FP problems in several situations by learning the underlying mapping function from the measurement to the solution, and it turns out that neural networks can indeed recover valuable sample images from static or temporal datasets in a short time period \cite{14,15}. However, the above applications can only reconstruct the amplitude portion of the objects. The ability to recover the lost phase component can provide valuable instructive information from another perspective. Inspired by \cite{16}, Zhang \emph{et al.} \cite{17} preprocessed a synthesized 2-channels input containing amplitude and phase, and successfully recovered the sample image via a multiscale deep residual network. Furthermore, a neural network was used to learn the design of an LED source pattern to reduce the acquisition cost \cite{18}. All the aforementioned neural networks are based on supervised data-driven architectures which need massive amounts of data to learn an underlying mapping. However, since general FP datasets are scarce and image capturing conditions are diverse, it is hard for these networks to be widely applied. Under these circumstances, physics-based neural networks have been employed to reduce the required amount of data by learning some specific scalar weights of an LED matrix, but it still requires a lot of training data \cite{19,20}.

In order to compensate for the dilemma of insufficient training data and make the framework available under various FP microscopy setups, we design an interpretable physics-guided neural network (PgNN) by modeling the Fourier ptychographic forward imaging process via the neural network. Besides, we also incorporate various optical and deep-learning tools into our model. With a combination of a physics-based network and an alternate updating mechanism, our model can unroll the intensity and phase images of a sample with both wide FOV and high resolution without the need for a ground truth and prior aberration knowledge, and the pupil function of the objective lens can be recovered simultaneously. Next, by introducing the total variation function on both the intensity and the phase components of the sample object, PgNN is able to obtain superior results. In addition, we incorporated the Zernike polynomial into our model to improve the modality of recovered aberration. Through our experiments, we demonstrate that PgNN works well in both simulated and experimental datasets. Specially, PgNN can outperform traditional FP methods when recovering a object in the complex domain from a high-defocus, high-exposure tissue section dataset.

In summary, our proposed model can obtain super-resolution results from a dataset captured from a single object without any supervision by designing a physics-guided framework and modeling targets as parameters of the network's hidden layers. In addition, we introduced optical theories and network tools to verify how these parameters can affect network performance. From the simulated and experimental results, we find that these methods do indeed improve the output performance and make the model more robust in extreme microscopy conditions.

This paper is structured as follows. In Section 2, we provide a brief review of developments in biomedical applications which utilize DNN to reduce human efforts. In Section 3, we discuss the fundamental principles and reconstruction procedure of FP. In Sections 4 and 5, we provide a detailed description of the development of PgNN, enhancing its performance step by step and validating it on both simulated and experimental datasets under variant acquisition conditions. Finally, we summarize the experimental results and discuss our on-going efforts in Sections 6 and 7.

%% file: 2.relatedwork.tex
\section{Related work}
Traditionally, to solve the non-linear inverse imaging problem, case-by-case analysis is necessary. By explicitly defining the type of problem and carefully engineering domain knowledge to design a dedicated imaging model, analytical methods can find optimized results \cite{8,22,23}. Unlike this sophisticated process, DNN does not benefit from such prior knowledge but utilizes structured frameworks to dig into the underlying mapping towards the inverse problem in massive datasets. Due to the infinite possibilities of DNN, more and more biomedical imaging problems are being solved with the neural network, such as computed tomography (CT) \cite{23}, image super-resolution \cite{24}, magnetic resonance imaging (MRI) \cite{25,26},  holography \cite{8,16,27,28} and cross-modality learning in CT/MRI  \cite{29,30}. DNN is even more widely used outside the scope of biomedical imaging applications.

However, generic datasets are always scarce due to patient privacy and the diversity of the acquisition environments. For these reasons, although data-driven models can always provide reliable solutions when there exists sufficient data, the problem of over-fitting or generalization has always plagued us in biomedical applications.

Some physics-based models have already been proposed to overcome this dilemma, making it possible to converge quickly with a small number of samples in signal processing and so on \cite{19,31,32}. Following this idea, we attempt to design a pure physics-guided model to address the limitation of insufficient training data and utilize empirical tools to further improve the performance of such a framework.

%% file: 3.background.tex
\begin{figure*}[b]
	\begin{center}
		\includegraphics[width=1\linewidth]{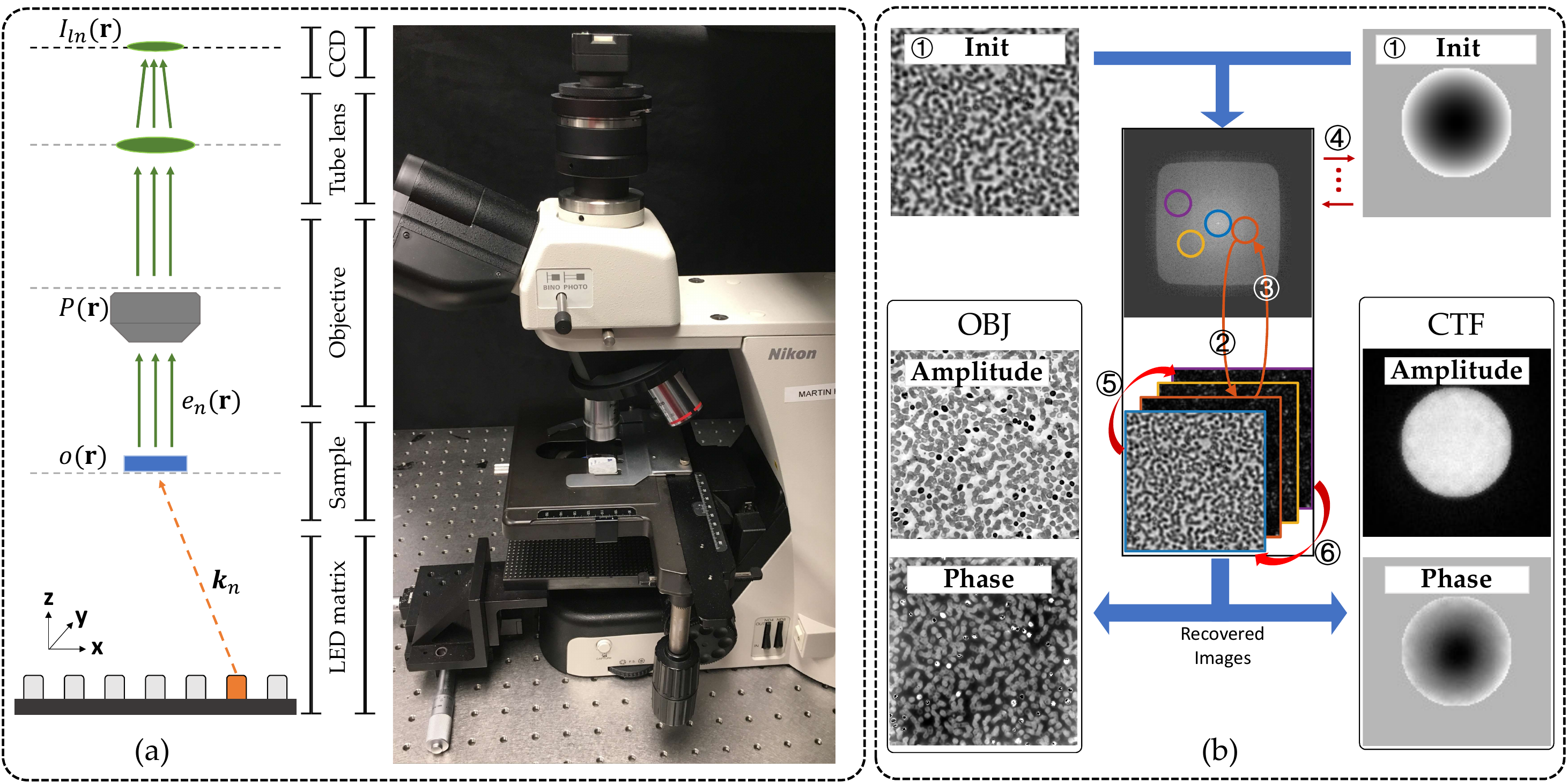}
	\end{center}
	\caption{Fundamental principles of Fourier ptychography. (a) The schematic diagram of the FP experimental setup and the physical comparison. (b) The iterative decomposition procedure embedded with the pupil function recovery process, where the sequential captured intensity images are used to recover the high-resolution sample image and CTF through steps 1-6. Step 1: initialize the high-resolution object and CTF. Step 2: extract a low-resolution aperture of the sample's Fourier spectrum corresponding to a unique wave vector. Step 3: replace the intensity component with the corresponding captured image in spatial domain while keeping the phase component unchanged to obtain the high-resolution aperture. Step 4: apply (6), (7) to remove the effect of optical aberration and thus correct two targets. Step 5: repeat steps 2-4 for other oblique wave vectors until all captured images are used. Step 6: repeat steps 2-5 several times until convergence.}
	\label{fig_1}
\end{figure*}

\section{Fourier Ptychographic Microscopy}
As a classic analytical method, the Fourier ptychography is mainly composed of the explicit forward imaging model and the decomposition procedure. Considering the generalized FP schematic diagram setup shown in Fig. \ref{fig_1}a, the sample is illuminated by an oblique plane wave from the LED matrix, and the exit wave is then captured by the camera through the objective lens. By sequentially lighting distinct LEDs on the matrix, FP can get a series of low-resolution intensity images to recover a high-resolution one.

In the forward procedure, we denote the thin sample as its transmission function $o(\textbf{r})$, where $\textbf{r}=(x, y)$ represents the 2D spatial coordinates with its Fourier expression as $\textbf{k}=(k_x,k_y)$. When illuminated by the ${\rm n}^{\rm th}$ oblique monochromatic LED, the reflected light wave can be denoted as $e_n(\textbf{r})=o(\textbf{r})\odot{\rm exp}(i\textbf{k}_{n})$, where `$\odot$' denotes the element-wise multiplication and $\textbf{\rm k}_{n}=(k_x^n,k_y^n)$ donates the ${\rm n}^{\rm th}$ wave vector corresponding to the angle of incident illumination. The final wave captured by the camera can be expressed as
\begin{equation}
I_{ln}(\textbf{r})=|e_n(\textbf{r})\ast P(\textbf{r})|^2=|o(\textbf{r})\ast\{{\rm exp}(i\textbf{k}_{n})\cdot P(\textbf{r})\}|^2,
\label{eq1}
\end{equation}
where `$\ast$' denotes convolution, $P(\textbf{r})$ denotes the PSF of the objective lens and $I_{ln}(\textbf{r})$ denotes the corresponding captured low-resolution intensity image. Here, we note that subscripts $l$, $h$ and $n$ denote low-resolution, high-resolution and sequence number, respectively. Since the convolution operation in spatial domain equals to the multiplication operation in Fourier domain, Equation (\ref{eq1}) can be further interpreted as
\begin{equation}
I_{ln}(\textbf{r})=|\mathcal{F}^{-1}\{O(\textbf{k})\odot C_n(\textbf{k})\}|^2,
\label{eq2}
\end{equation}
where `$\mathcal{F}$' refers to Fourier transformation, $C_n(\textbf{k})=\mathcal{F}\{{\rm exp}(i\textbf{k}_{n})\cdot P(\textbf{r})\}=C(\textbf{k}-\textbf{k}_n)$ denotes the corresponding Fourier spectrum of the ${\rm n}^{\rm th}$ PSF, and $O(\textbf{k})$ denotes the Fourier transformation of $o(\textbf{r})$. The standard formulation of the coherent transfer function (CTF) can be expressed as
\begin{equation}
C(\textbf{k})=step\{(k_x^2+k_y^2)<({\rm NA}\cdot k_0)^2\},
\label{eq3}
\end{equation}
where `$step$' refers to the step function, NA characterizes the range of angles over which the system can accept light, $k_0=2\pi/\lambda$ and $\lambda$ is the illumination wavelength.

For diffraction-limited imaging, the CTF is a strict circular passband located at the center of the Fourier spectrum, and $C_n(\textbf{\rm k})$ is equal to the shift of the CTF center to the location $\textbf{k}_{n}$. As such, the high frequency information outside the original CTF is shifted into the passband of the low-NA objective lens and can be captured by the camera. Then, FP can synthesize these images with different spectral information to estimate a high-resolution result. Generally, FP reconstructs the high-resolution image through a phase retrieval theory called alternating projection (AP), which iteratively updates distinct circular subregion by keeping its spatial phase unchanged and replacing its spatial amplitude with the square root of corresponding captured intensity image, to recover the phase component. One classic implementation of AP can be formulated as
\begin{equation}
\varPhi_{ln}(\textbf{r})=\mathcal{F}^{-1}\{\varphi_{ln}(\textbf{k})\},
\label{eq4}
\end{equation}
\begin{equation}
\varPhi_{hn}(\textbf{r})=\sqrt{I_{ln}(\textbf{r})}\odot\varPhi_{ln}(\textbf{r})/|\varPhi_{ln}(\textbf{r})|,
\label{eq5}
\end{equation}
where $\varphi_{ln}(\textbf{k})=O(\textbf{k})\odot C_n(\textbf{k})$ denotes the specific subregion of the sample's Fourier spectrum under ${\rm n}^{\rm th}$ illumination, and $\varPhi_{hn}(\textbf{r})$ denotes the updated high-resolution image in the spatial domain.

The premise of (\ref{eq4}), (\ref{eq5}) is that the sample should be strictly placed on the focal plane, which requires careful calibration in  practical applications \cite{33}. Meanwhile, as the low-NA objective lens is not originally designed for high resolution imaging applications, the optical aberration becomes the limiting factor when increasing image resolution~\cite{2}. As such, FP always embeds an extra method after AP to compensate for this interference. The supplementary decomposition process is usually composed of two parts. Firstly, the object reconstruction process can be expressed as
\setlength{\arraycolsep}{0.14em}
\begin{eqnarray}
\varphi'_{hn}(\textbf{k})&{}={}&\varphi_{hn}(\textbf{k})+\alpha\cdot C_m^\ast(\textbf{k})/|C_m(\textbf{k})|^2_{max}\nonumber\\
&&{\odot}[\varphi_{hn}(\textbf{k})-\varphi_{ln}(\textbf{k})].
\label{eq6}
\end{eqnarray}
This function is used to update the sample spectrum under the current aberration function $C_m(\textbf{k})$ and the $\rm n^{\rm th}$ wave vector $\textbf{k}_n$, where $\varphi_{hn}(\textbf{k})=\mathcal{F}\{{\varPhi_{hn}}\}$ denotes the updated high-resolution subregion corresponding to $\varphi_{ln}(\textbf{k})$, and $\varphi'_{hn}(\textbf{k})$ is corrected from the difference of those two exit waves during AP.

Next, with a similar form, the aberration update process for CTF can be expressed as
\setlength{\arraycolsep}{0.14em}
\begin{eqnarray}
C_{m+1}(\textbf{k})&{}=&{}C_m(\textbf{k})+\beta\cdot\varphi_{hn}^\ast(\textbf{k})/|\varphi_{hn}(\textbf{k})|^2_{max}\nonumber\\
&&\odot[\varphi_{hn}(\textbf{k})-\varphi_{ln}(\textbf{k})],
\label{eq7}
\end{eqnarray}
where the subscript `$m$' is used to represent the current state of the CTF. In addition, subscript `$max$' refers to the maximum function and the superscript `$^\ast$' denotes the conjugation. The constants $\alpha$ and $\beta$ represent the step size of the optimization, which usually take 1 for majority cases. We should note that (\ref{eq6}), (\ref{eq7}) are usually used to correct the effects of the optical aberration CTF after the phase retrieval process defined in (\ref{eq4}), (\ref{eq5}). 

Following the above principles, the modified FP decomposition procedure is shown in Fig. \ref{fig_1}b, where steps 1-6 are used to illustrate the overall process. In the first step, FP initializes the object and CTF, where the up-sampled captured intensity image under the central LED with zero phase is a good starting point for the object and the step function defined in (\ref{eq3}) is a general initial guess for the CTF (a guess of a tiny defocus distance also works). Following the AP defined in (\ref{eq4}), (\ref{eq5}), FP selects a circular subregion of the object under a particular illumination angle in Fourier domain (step 2) and updates the frequency information by replacing the spatial amplitude component with the corresponding captured image (step 3). Then, to eliminate the artifacts caused by the unknown CTF, step 4 applies (\ref{eq6}),~(\ref{eq7}) to further update these two targets. In step 5, steps 2-4 are repeated for other illuminations until the whole frequency information is updated. It is critical that each extracted subregion must overlap with adjacent subregions to assure convergence. At the end, FP repeats the above procedures to get self-consistent solutions for the sample and CTF.

%% file: 4.method.tex
\begin{figure*}[!t]
	\centering
	\includegraphics[width=0.94\linewidth]{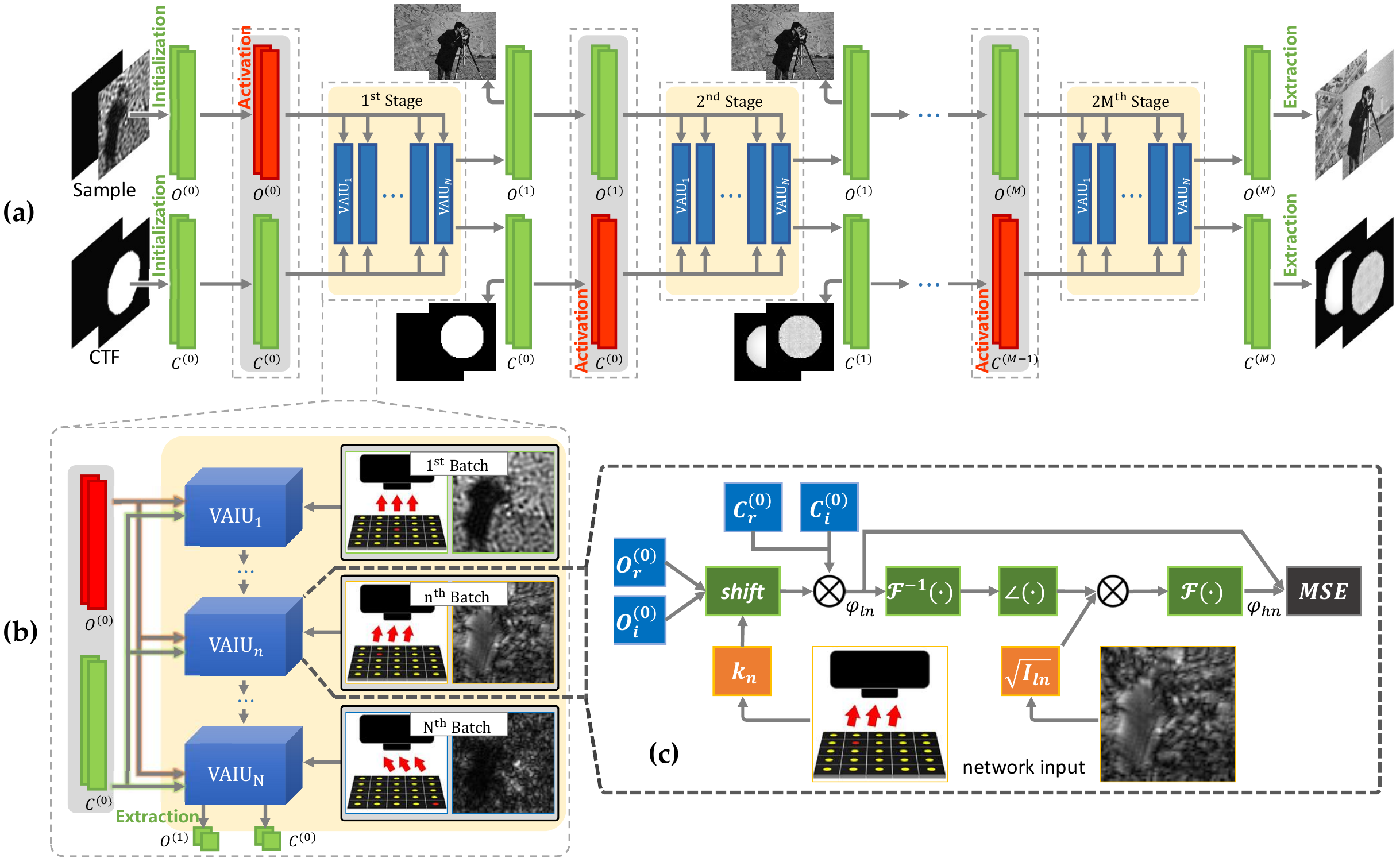}
	\caption{Illustration of the PgNN framework. Specifically, the targets needed to be unrolled are treated as learnable hidden layer parameters and optimized through back propagation. (a) Schematic of pipeline procedure with the alternate updating mechanism
. PgNN is composed of 2M stages, and the update target differs in each stage. After initializing the sample and CTF as $O^{(0)}$ and $C^{(0)}$ in the complex domain, the sample parameters $O^{(0)}$ are activated in the first stage and updated by VAIUs. Then, $C^{(0)}$ is updated in the second stage and the reconstructed targets are extracted after the final stage. For better understanding, the amplitude and phase components of the targets at different stages are shown in corresponding positions. (b) Schematic of the first stage, where each VAIU shares the same network structure and learnable parameters $O^{(0)}$, and has its own dataset input. We should note that each VAIU is essentially equivalent to a forward propagation process of our neural network under different batch inputs, and the optimized results will be extracted from corresponding hidden layers in (c). (c) The basic framework of VAIU corresponding to (\ref{eq4.1.1}),~(\ref{eq4.1.2}),~(\ref{eq4.1.3}).}
	\label{fig_2}
\end{figure*}

\section{Method}
\subsection{The Basic Framework of PgNN}
Introducing professional knowledge into a neural network model is equivalent to providing a strong prior condition for datasets, which is helpful for the network to better learn underlying mappings within limited data. Following this idea, we model the forward imaging process via a neural network called physics-guided neural network (PgNN), which could quickly reconstruct the sample image from captured intensity images illuminated under various angles. To enhance the capability to distinguish characteristics of the sample and CTF respectively, we mainly model the framework in the Fourier domain and transfer the whole FP forward imaging process into the neural network
\begin{equation}
\varphi_{ln}(\textbf{k})=O(\textbf{k}+\textbf{k}_n)\odot C(\textbf{k}),
\label{eq4.1.1}
\end{equation}
\begin{equation}
\varphi_{hn}(\textbf{k})=\mathcal{F}\{\sqrt{I_{ln}(\textbf{r})}\odot\angle\{{\mathcal{F}}^{-1}\{\varphi_{ln}(\textbf{k})\}\}\},
\label{eq4.1.2}
\end{equation}
where $\varphi_{ln}$, $\varphi_{hn}$ represent the original and updated subregion in Fourier domain during the AP and `$\angle$' denotes the phase component. 

Fig. \ref{fig_2} illustrates the detailed pipeline procedure and basic framework of PgNN. Specifically, PgNN is composed of several training stages, and each stage consists of different units. As shown in Fig. \ref{fig_2}b, each unit has its own physical meaning, which strictly corresponds to one forward imaging model under the unique batch input of the plane wave incident $\textbf{k}_n$ and the captured intensity image $I_{ln}(\textbf{r})$. Here, we name these units as varying angle illumination units (VAIUs). By sequentially inputting the parameters of sample and CTF into these VAIUs, the network can extract the high-resolution sample image and optical aberration CTF in the complex domain from the feature maps of corresponding hidden layers. Without confusion, each VAIU essentially corresponds to a forward propagation process of neural networks.

The basic framework of VAIU is shown in Fig. \ref{fig_2}c. This unit takes the wave vector $\textbf{k}_n$ and the square root of captured intensity image $I_{ln}$ as the network inputs, and the object image and CTF are then naturally treated as two-channel (the real and imaginary parts) learnable feature maps of hidden layers to participate in the training procedure. Following the forward imaging model defined in (\ref{eq4.1.1}),~(\ref{eq4.1.2}), the target object $O(\textbf{k})$ is forward propagated through the well-designed framework to generate a subregion $\varphi_{ln}(\textbf{k})$ of the Fourier spectrum under the conditions of $\textbf{k}_n$ and the other target ${C}(\textbf{k})$. Then, the low-resolution spectrum $\varphi_{ln}(\textbf{k})$ is transmitted via the AP with $I_{ln}(\textbf{r})$, and an updated high-resolution spectrum $\varphi_{hn}(\textbf{k})$ is generated. 

According to the phase retrieval mechanism, the two exit wave series of predicted spectra and updated spectra should contain the same frequency information when converged. Therefore, the differences between these spectra are back propagated to optimize the whole network, and the model can simultaneously recover the object and the pupil function without supervision in ideal situation. The basic loss function used in PgNN can be expressed as
\setlength{\arraycolsep}{0.14em}
\begin{eqnarray}
loss&{}=&{}\Sigma^{N}_{n=0}\Vert\varphi_{hn}(\textbf{k})-\varphi_{ln}(\textbf{k})\Vert^2_2,
\label{eq4.1.3}
\end{eqnarray}
where L2-norm is used to measure the differences and $N$ denotes the number of varying illumination angles.

In addition, we should note that this network is modeled in the complex field and all the layers are defined with the real and the imaginary parts. For example, we can rewrite (\ref{eq4.1.1}) in the complex domain as
\begin{eqnarray}
\varphi_{ln}(\textbf{k})&{}=&{}\{O_r(\textbf{k}+\textbf{k}_n)+i\cdot O_i(\textbf{k}+\textbf{k}_n)\}\nonumber\\
&&\odot \{C_r(\textbf{k})+i\cdot C_i(\textbf{k})\}\\
&{}=&{}(O_r\odot C_r-O_i\odot C_i)+i\cdot(O_r\odot C_i+O_i\odot C_r)\nonumber,
\label{eq4.1.4}
\end{eqnarray}
where the subscripts $r$ and $i$ denote the real and imaginary parts, respectively. After traversing the whole dataset, we can extract parameters of hidden layers to get the optimized sample and CTF.

\subsection{Alternate Updating Mechanism}
Theoretically, we can obtain the high-resolution sample and CTF by repeating the iteration procedure within several stages. However, since the inputs of the PgNN only contains finite images (corresponding to different illumination angles of a single object) and each image contains redundant information due to the frequency overlap, the sample and CTF will interfere with each other in backpropagation and the reconstructed image will be highly blurred. In order to eliminate this dilemma, we devise an alternate updating mechanism
 to ensure good results.

As shown in Fig. \ref{fig_2}a, unlike the general consecutive iterative process, PgNN is composed of 2M stages and the updating objectives differ in different stages. In the first stage, we believe the CTF is closer to the practical ground truth than the sample, and PgNN will focus on updating the sample parameters $O^{(0)}$ while keeping the CTF unchanged.  Then, we will update the CTF $C^{(0)}$ by keeping $O^{(1)}$ unchanged, since the modified sample $O^{(1)}$ after previous updating will contain more detailed and realistic information. As the network parameters are alternately updated, the high-resolution object and CTF will gradually converge to the optimal point. From the corresponding amplitude and phase components shown in Fig. \ref{fig_2}a, we can see the obvious resolution improvement in the first two stages, and obtain reconstructed results with high quality after the last stage.

Another benefit to incorporating the alternate updating process is that this mechanism can help to update the parameters in a more efficient way and make the model converge quickly by changing the backward information flow \cite{35}. In our experiments, we typically take five epochs or fewer for a single stage to make a quick optimization.

\subsection{The Enhanced Loss Function}
One of the strengths of neural networks lies in the diversity of their powerful loss functions. Since different loss functions lead to different backward gradients, the models can learn the underlying information under various gradient descents. Here, we apply the total variation (TV) regularization into our model to improve the performance of the network.

TV is a widely used loss function in the field of signal processing and image super-resolution \cite{36,37}. By calculating the integral of the absolute gradient, it makes an evaluation of the degree to which the image is disturbed by noise. Since the TV loss encourages spatial smoothness in generated images, we incorporated it in our method to effectively eliminate artifacts and make the images more realistic. 

For our FP application, we attempted to reconstruct the object images composed of both amplitude and phase components. Thus, we expand the basic loss function (\ref{eq4.1.3}) with TV as follows,
\begin{eqnarray}
loss&{}=&{}\Vert\varphi_{hn}(\textbf{k})-\varphi_{ln}(\textbf{k})\Vert^2_2+\alpha_1\cdot TV\{|\varPhi_{hn}(\textbf{r})|\}\nonumber\\
&&{+}\alpha_2\cdot TV\{\angle\varPhi_{hn}(\textbf{r})\},
\label{eq4.3.1}
\end{eqnarray}
%\begin{equation}
%	TV_1\{\varPhi_{hn}(\textbf{r})\}=TV\{|\varPhi_{hn}(\textbf{r})|\},
%\end{equation}
%\begin{equation}
%	TV_2\{\varPhi_{hn}(\textbf{r})\}=TV\{\angle\varPhi_{hn}(\textbf{r})\},
%\end{equation}
\begin{equation}
TV\{o(\textbf{r})\}=\sum(|o_{x+1,y}-o_{x,y}|^2+|o_{x,y+1}-o_{x,y}|^2)^{\eta/2},
\label{eq4.3.2}
\end{equation}
where `$|\cdot|$' and `$\angle$' represent the amplitude component and the phase component of the object. Equation (\ref{eq4.3.1}) is composed of three parts, where $\Vert\varphi_{hn}(\textbf{k})-\varphi_{ln}(\textbf{k})\Vert^2_2$ denotes the deviation during the AP, $TV\{|\varPhi_{hn}(\textbf{r})|\}$ and $TV\{\angle\varPhi_{hn}(\textbf{r})\}$ represent the smoothness of the amplitude and phase of the updated high-resolution image in the spatial domain. Parameters $\alpha_1$ and $\alpha_2$ denote the coefficients of two TV terms respectively. The larger the value, the smoother the reconstructed image. The standard formulation of the TV function is expressed as (\ref{eq4.3.2}), where $o(\textbf{r})$ is used as a template, $\eta$ represents the power series of TV and $\eta=1$ is used in our experiments.

This loss function is then optimized via the backward stochastic gradient descent, and it is able to adjust the network output by applying different parameter combinations.

\begin{figure}[!t]
	\centering
	\includegraphics[width=0.9\linewidth]{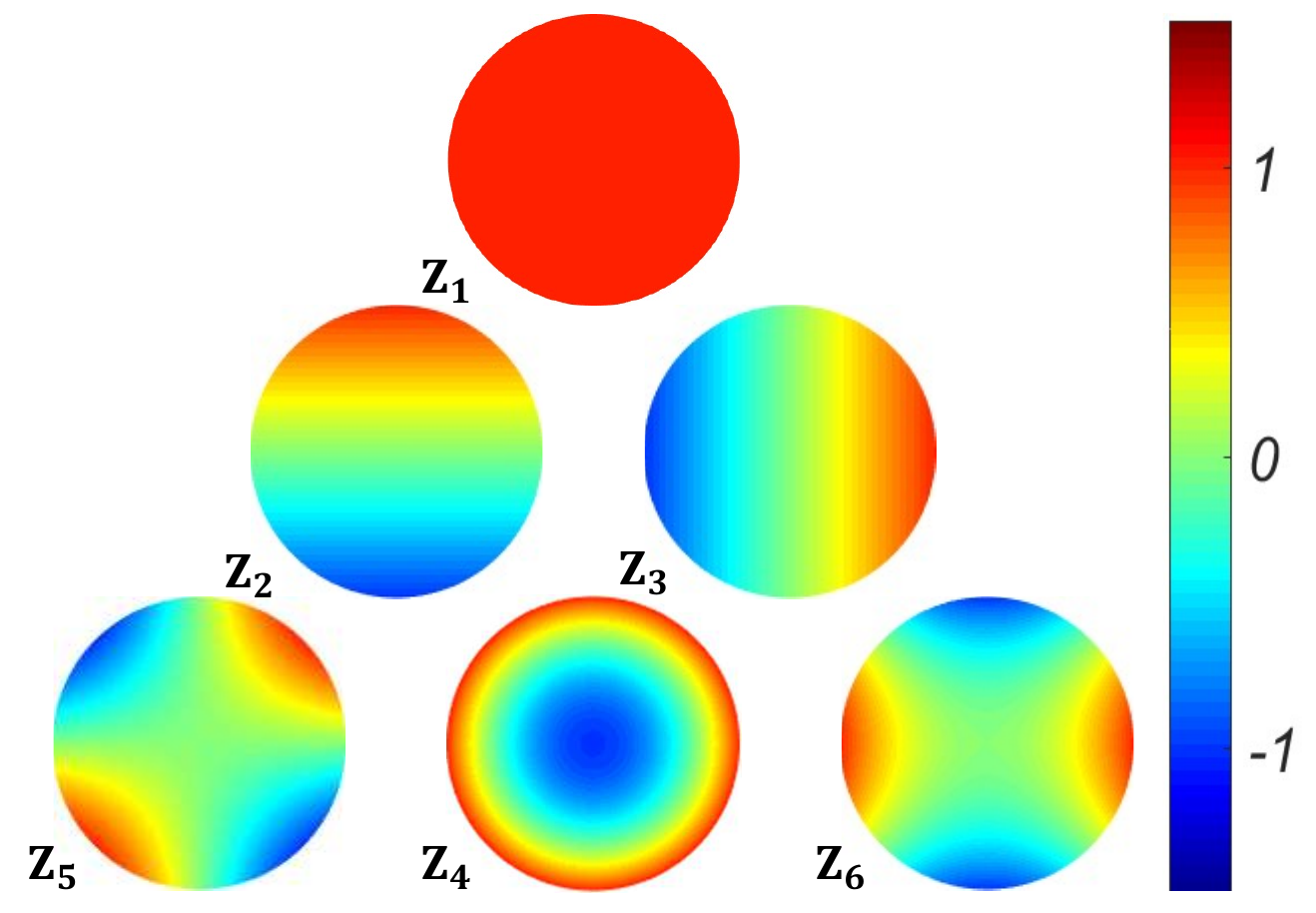}
	\caption{The color bar of the first six Zernike order terms. ${\rm Z}_1$ represents the bias term. The second and third modes denote the tilt in the x or y axis. ${\rm Z}_4$ is the defocusing. The last two modes are the primary astigmatism in the x or y axis.}
	\label{fig_3}
\end{figure}

\begin{figure*}[b]
	\begin{center}
		\includegraphics[width=0.925\textwidth]{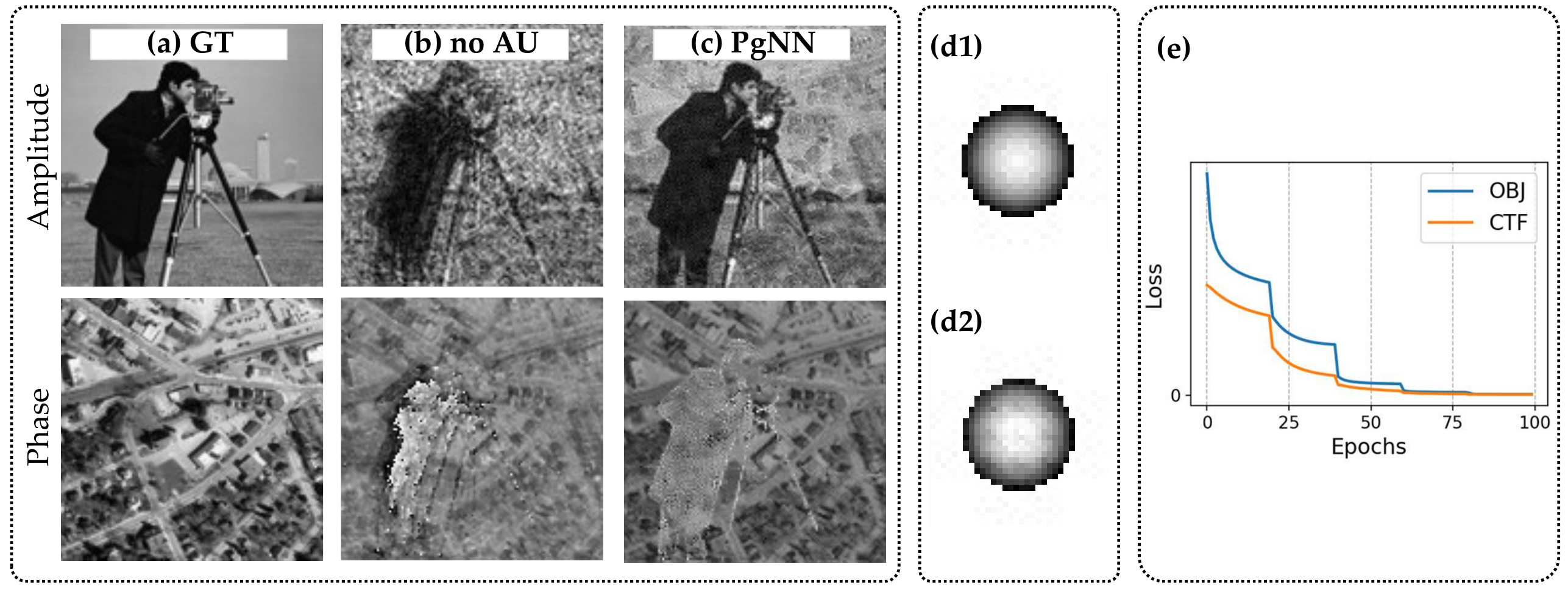}
	\end{center}
	\caption{Comparison of recovered images on simulated dataset. (a) Ground truth of amplitude and phase components. (b), (c) Recovered sample images from limited PgNN where both the TV function and Zernike mode are not activated, and alternate updating (AU) mechanism is not activated in (b). (d1) Ground truth of CTF, caused by defocus distance ${\rm z}=50 {\mu m}$. (d2) Recovered CTF from (c). (d1), (d2) denote phase components of CTF, and amplitude components are not presented. (e) Training curve for object and CTF only with the alternate updating mechanism.}
	\label{fig_4}
\end{figure*}

\subsection{Improved Modality of Optical Aberration}
When observing the reconstructed results, we find it would be hard and ambiguous to recover a high quality CTF, since the reconstructed phase component contains many artifacts. Although the application of deep learning tools to improve the performance of neural networks is a preferred and effective method, it would be much desirable to update the aberration in a more efficient and effective optical manner.

Here, we incorporated the Zernike mode into our proposed network to further improve the recovery accuracy of the CTF, due to its powerful ability to describe wavefront characteristics in optical imaging \cite{38}. The Zernike modes are composed of a series of polynomials defined orthogonal to each other within a unit circle, which are consistent with the systematic aberration forms observed in microscopy. As a power series expansion, each order term of the Zernike polynomial has its unique corresponding physical meaning. Fig. \ref{fig_3} shows the color bar of the first six Zernike modes, where ${\rm Z}_1$ denotes the constant term in the optical aberration, which makes no contribution towards the FP decomposition procedure, and ${\rm Z}_4$ represents the defocus pattern, which is quite common in captured datasets \cite{1,2,33}.

In traditional methods, the CTF is always updated as a whole. When directly inherited in PgNN, it would take large number of stages to converge. Nevertheless, if we apply the Zernike function to change the form of hidden layers of the CTF, the parameters that need to be compensated for the aberration are reduced from $o({n}^{\rm 2})$ to the constant number of Zernike modes that need to be fitted. The new modality of the phase component of the CTF is updated as
\begin{equation}
\angle C (\textbf{k})=\sum\nolimits_{l=1}^{L}c_l\cdot Z_l(\textbf{k}),
\label{eq4.4.1}
\end{equation}
where $L$ denotes the number of Zernike modes, and $c_l~\in~{R}$ is the coefficient of each Zernike polynomial $Z_l(\textbf{k})$. In general, the first nine modes can already fit the common aberration in microscopy. In other words, we only need to train nine parameters to achieve the same performance as that of its entirety. As for the CTF intensity component, it will still be updated as a whole. As such, $C(\textbf{k})=|C(\textbf{k})|\cdot exp\{i\angle C(\textbf{k})\}$ is the final form we used in PgNN.

The additional benefits of applying Zernike modes are as follows. The capacity of our datasets is very small, and the previous strategy has the problems of insufficient training and difficulty in CTF reconstruction. When training finite coefficients on the optical aberration, PgNN can focus on the most important aberration patterns, such as defocusing, and recover a practical CTF, which can help to improve network performance.

\subsection{FP Reconstruction Procedure with PgNN}
By combining the above mechanisms, PgNN can effectively reconstruct the objects and CTFs with high quality and the whole procedure can be divided into three parts (Fig. \ref{fig_2}a).

In the first initialization section, we will initialize the parameters required by the network in a novel way. The dataset fed into the network contains a group of low-resolution intensity images captured from a single sample illuminated under various angles. We should note that as the dataset contains the high frequency information necessary to recover the object, some useful methods like data augmentation were not applied. Additionally, since the physics-guided network is quite different from other data-driven models, the traditional sophisticated weight initialization method is no longer suitable in this situation. Consistent with traditional methods, the up-sampled central intensity image and the standard format of CTF (\ref{eq3}) with zero phase component are used as the initial guesses.

Next, the object and CTF are modeled as learnable weights of hidden layers according to (\ref{eq4.1.1}), (\ref{eq4.1.2}) and (\ref{eq4.4.1}). Since the TV coefficients $\alpha_1$, $\alpha_2$ are usually negligible, the loss function (\ref{eq4.4.1}) is mainly determined by the deviation extracted from the two exit waves during AP, and the network output approaches zero when converges. In this way, PgNN can obtain the results without any supervision, and can adjust the imaging performance by choosing different combinations of TV coefficients. Considering multiple epochs as a stage, the network can converge in a short number of stages by updating these two targets alternately via backpropagation.

At the output stage, we can obtain the wide FOV, high-resolution object with unrolled CTF simultaneously by extracting the optimized parameters of hidden layers.

%% file: 5.experiment.tex
%\begin{figure}[tp]
%	\centering
%	\includegraphics[width=1.0\linewidth]{Fig7}
%	\caption{Reconstructions on a tissue section stained by immunohistochemistry methodology. (a), (b), (c) show the recovered complex object images from ePIE, PbNN and PgNN. The first row denotes the recovered color intensity images and the second denotes the phase components.}
%	\label{fig_7}
%\end{figure}

\section{Experiment}
We validated our PgNN on both simulated and experimental datasets and compared our quadruple sampling results with the classic analytical method ePIE \cite{2}, which plays the role of ground truth in many data-driven FP models. Because there are no generic automatic indicators for biomedical images, we will take the smoothness and detailed information such as cellular structures as the evaluation standards.

\subsection{System Setup}
We used a programmable LED matrix placed 9 cm below the sample plane for illumination, with the step size for adjacent incident wave vectors $\textbf{k}_n$ being 0.05 in both $k_{x}$ and $k_{y}$. We then introduced a 2X (the magnification factor is 2), 0.1 NA objective lens with a 200 mm tube lens to build a FP microscopy platform and utilized a 5-megapixel camera with a pixel size of 3.45 $\mu m$ to sequentially capture low-resolution intensity images. The whole setup and its schematic diagram are shown in Fig. \ref{fig_1}a.

For the generation of simulated datasets, we first chose two images as the amplitude and phase components of the ground truth in the spatial domain. Then, 225 low-resolution subregions in the sample's spectrum were generated corresponding to the 15 by 15 LEDs in the illumination matrix according to the physical model defined in (\ref{eq2}), and the intensity components with a size of $32\times32$ obtained by the inverse Fourier transform were used as simulation data. To be more authentic, we additionally assumed these images were captured with a defocus distance of 50 $\mu m$ which would raise the difficulty of achieving convergence.

For the acquisition of experimental datasets, we collected images under two image capturing conditions. First, we obtained two tissue section datasets where the samples were well placed on the focal plane. Under this circumstance, the CTF could hardly exert influence on the sample reconstruction and we would compare images recovered from PgNN and ePIE. The corresponding image size was $64\times64$. Then, we designed an extreme but valuable extension that we placed the sample in a high-defocus, high-exposure condition with the captured image size of $128\times128$ under monochromatic light source. In this extension, some captured images were so overexposed that their corresponding frequency information were totally lost.

For the settings of PgNN, we alternately updated the sample and CTF generally for 10 stages, and 5 epochs for each stage in experimental datasets. As for simulated datasets, such hyperparameters would be set larger since the image size was smaller. The input image size could be arbitrary, and the up-sampled output image size was 4 times larger than the input. Furthermore, we applied an Adam optimizer to train PgNN.

\begin{figure*}[!b]
	\begin{center}
		\includegraphics[width=0.8\textwidth]{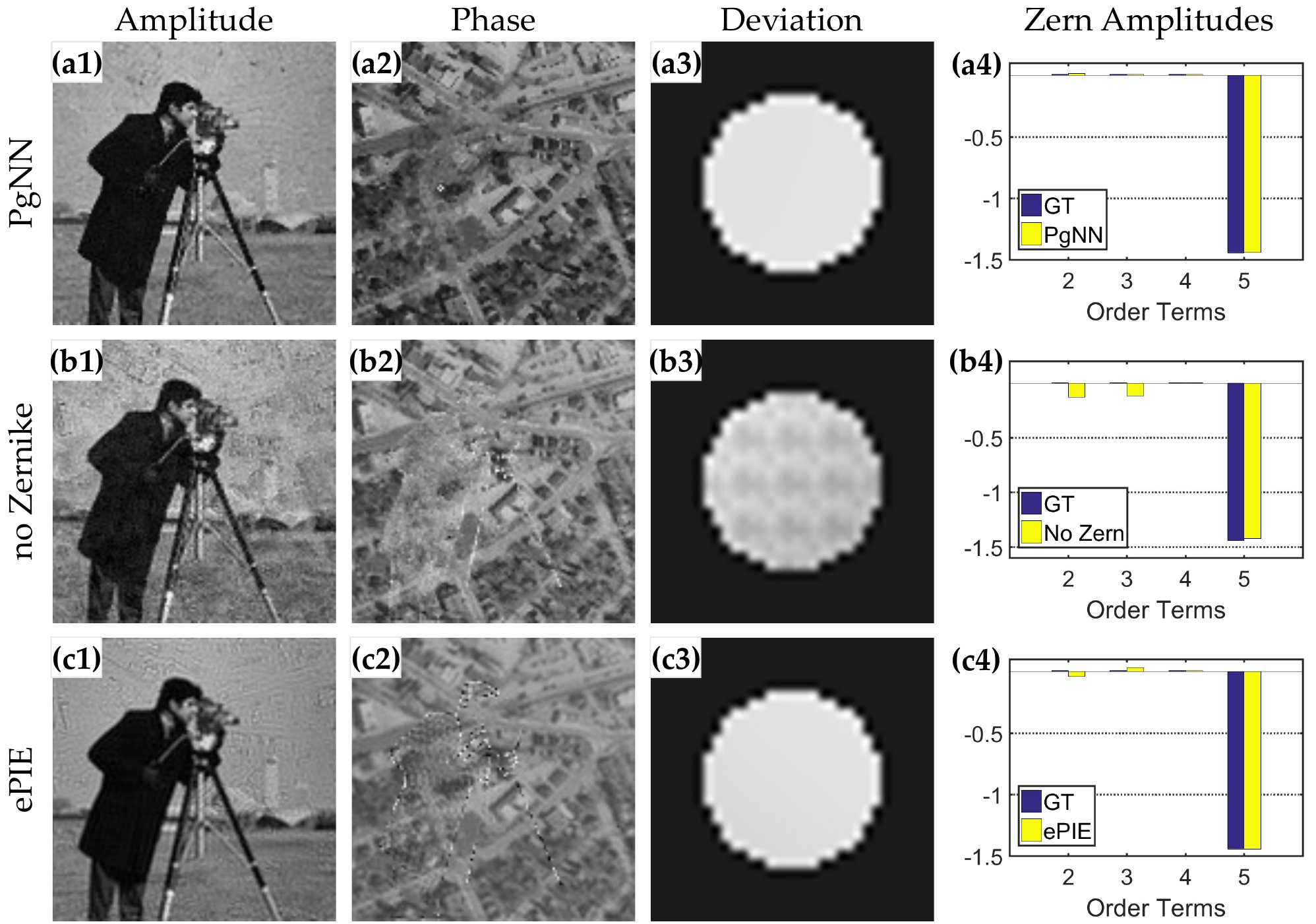}
	\end{center}
	\caption{Comparison of recovered sample images and CTFs. (a), (b) Results from PgNN where the Zernike mode is activated and not. (c) Results from ePIE. The first two columns show the sample images, and the third represents deviations between reconstructed and actual CTF. Histograms of some Zernike polynomial coefficients $c_l$ are shown in the last column.}
	\label{fig_5}
\end{figure*}

\begin{figure*}[t]
	\begin{center}
		\includegraphics[width=0.85\linewidth]{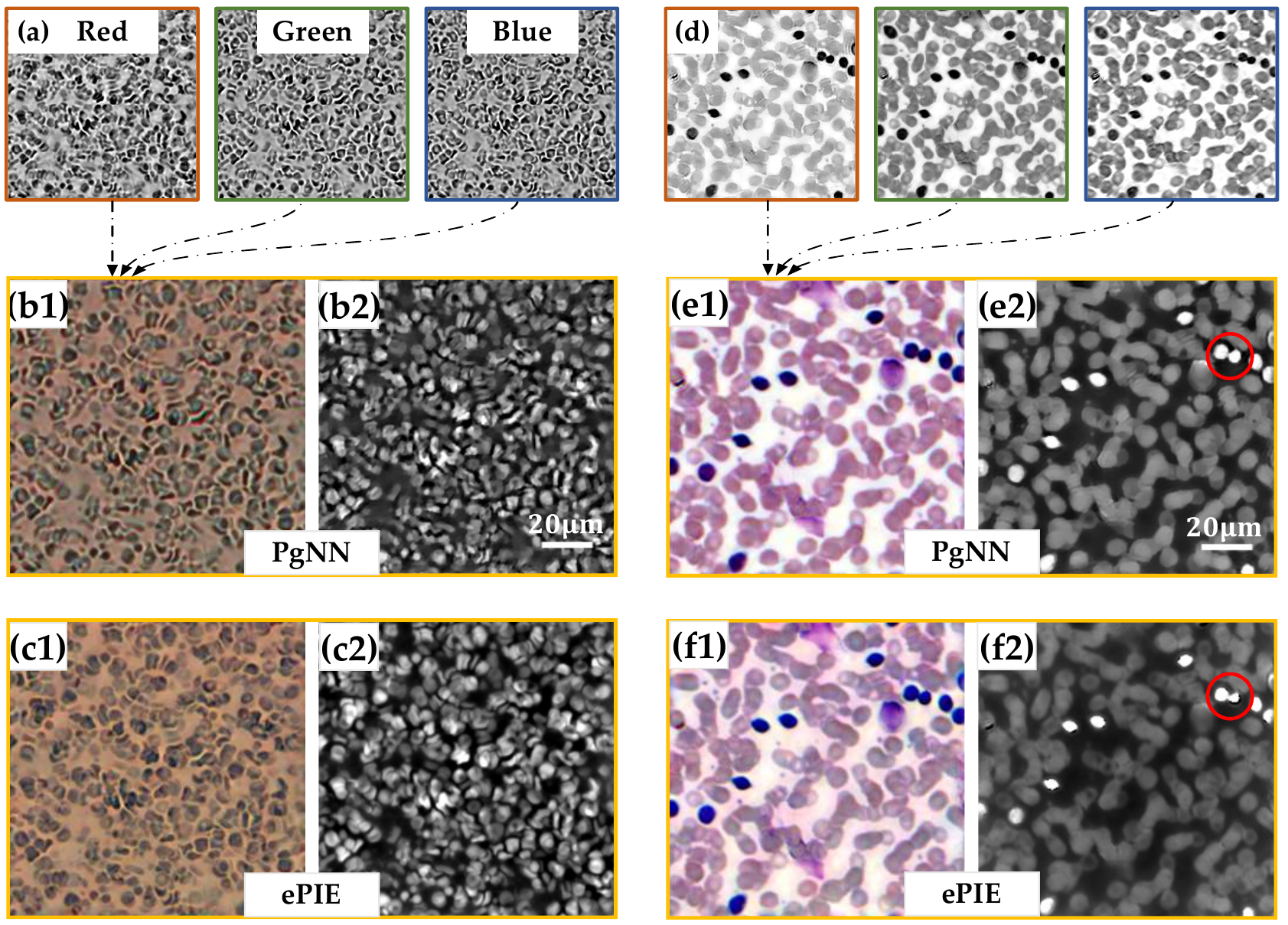}
	\end{center}
	\caption{Reconstruction of two datasets in low-defocus condition. (a) Recovered amplitudes at 632 nm, 532 nm and 470 nm wavelengths from PgNN with corresponding color borders. (b), (c) The combined color amplitude and phase images of a tissue section stained by immunohistochemistry methodology from PgNN and ePIE. (d), (e), (f) Corresponding results of blood cells towards (a), (b), (c), respectively. In addition, some raw captured intensity images of the blood smear dataset are shown in Fig. \ref{fig_1}b.}
	\label{fig_6}
\end{figure*}

\subsection{Results on Simulated Datasets}
The ground truths are shown in Fig. \ref{fig_4}a and d1, where the former denotes the amplitude and phase components of the sample of which image size is $128\times128$, and the latter represents the corresponding optical aberration in a defocus distance of 50 $\mu m$. The goal of PgNN is to unroll those features from $225\times32\times32$ intensity images. 

\textbf{Alternate updating process.} In Fig.~\ref{fig_4}c, we show the reconstructed results from incomplete PgNN where the TV loss and Zernike mode are not activated. Meanwhile, we additionally forced PgNN to continuously update the sample while keeping the CTF unchanged, and the corresponding results are shown in Fig.~\ref{fig_4}b. Apparently, the modified reconstruction (Fig. \ref{fig_4}c) contains fewer artifacts and achieves higher fidelity compared to the results shown in Fig. \ref{fig_4}b, owning to the reconstruction of the CTF (Fig.~\ref{fig_4}d2). As to the training curve corresponding to Fig.~\ref{fig_4}c, it is also obvious that the introduction of alternate updating principle indeed leads to a better result by the significant margins shown in Fig.~\ref{fig_4}e. 

We can demonstrate that the alternate updating mechanism raises the framework efficiency and accelerates convergence by controlling the backward dataflow. However, this incomplete PgNN has limitations in that the network cannot work well when the data is captured under high defocus distances like 50~$\mu m$. In this distance, the recovered aberration contains unexpected periodic speckles caused by incorrect updating sequences and are difficult to correct. To solve these problems, additional deep-learning and optical principles were incorporated.

\textbf{Zernike polynomial function.} As shown in Fig.~\ref{fig_5}, the greatest contribution made by the Zernike mode is the quality correction of the systematic aberration. Observing histograms shown in the last column of Fig.~\ref{fig_5}, it is found that the PgNN with a Zernike polynomial can precisely fit the polynomial coefficients (Fig.~\ref{fig_5}a4) while the one without that cannot restore these parameters (Fig.~\ref{fig_5}b4). More specifically, the biased coefficients in Zernike modes 2-3 cause speckles in the reconstructed optical aberration shown in Fig.~\ref{fig_5}b3. As reflected in the reconstructed images, the Zernike mode helps PgNN to achieve higher quality (Fig.~\ref{fig_5}a1,a2). In addition, as shown in Fig.~\ref{fig_5}c4, we can see that the coefficient amplitudes obtained by decomposing the reconstructed CTF from ePIE are different from the ground truth in Zernike modes 2-3. This disadvantage will hinder the high-quality reconstruction of sample images shown in Fig.~\ref{fig_5}c1,c2. Such a phenomenon also proves that taking the optical aberration as a power series expansion is better for learning relative features than taking it as a whole.

\textbf{Total variation loss.} By introducing the alternate updating mechanism and Zernike mode to the basic framework of PgNN, we built a generic interpretable neural network framework for FP reconstructions. Here, we want to further eliminate the interference between amplitude and phase components in reconstructed images, and TV loss is used to handle this problem by making a trade-off between these two targets. 

By adjusting the coefficient combination of two TV terms $\alpha_1$ and $\alpha_2$, PgNN can obtain the desired result and thus effectively eliminate image artifacts. Comparing the reconstructed amplitudes shown in Fig.~\ref{fig_4}c and Fig.~\ref{fig_5}b1 where the latter activates the TV function, it is obvious that these two supplementary losses greatly reduce the background noise caused by the phase image (Fig.~\ref{fig_5}b2). With a similar conclusion, the incorporated TV losses make the phase component smoother (Fig.~\ref{fig_5}b2) than without TV (Fig.~\ref{fig_4}c). In addition, when we activate the Zernike mode and compare our results (Fig.~\ref{fig_5}a) with those from ePIE (Fig.~\ref{fig_5}c), we clearly find that our reconstructed phase (Fig.~\ref{fig_5}a2) completely eliminates the amplitude interference while ePIE (Fig.~\ref{fig_5}c2) cannot. We can also see the performance improvement in amplitude components. As such, we demonstrate that the TV regularization can indeed improve the quality of reconstructed image by eliminating interference fringes.

As an additional supplement, the results from neural networks, such as PgNN, and traditional methods, such as ePIE, do differ from each other, and neural networks provide a new perspective for the FP reconstruction.

\subsection{Results on Experimental Datasets}
In this session, we will focus on the comparison of PgNN and traditional ePIE over various experimental datasets. Furthermore, we design a high-defocus experiment to validate that our method is strongly robust under extreme conditions.

\textbf{Normal condition.} We first test our framework over two general datasets where the samples are placed on the focal plane. Fig.~\ref{fig_6} shows the reconstructions of a tissue section stained by immunohistochemistry methodology and the blood smear. In Fig.~\ref{fig_6}a, we show the three recovered images of the tissue section illuminated by red, green and blue monochromatic LEDs. By combining the three images together, the unrolled color image is shown in Fig.~\ref{fig_6}b1 with its phase component shown in Fig.~\ref{fig_6}b2. The corresponding results from ePIE are shown in Fig.~\ref{fig_6}c1,c2. The tissue section dataset is carefully captured so that we could ensure the defocus distance is within a reasonable range. Therefore, the optical aberration could be ignored during the reconstruction, and we find that both PgNN and ePIE can obtain high-resolution images with high quality. 

Besides, we test the two methods in another blood smear dataset and draw the same conclusion. The recovered sample images from ePIE (Fig.~\ref{fig_6}f) and those from PgNN (Fig.~\ref{fig_6}e) share highly similar detailed information such as cellular structures and phase distribution, and we can clearly see the performance improvement from these reconstructions (some raw images are given in Fig.~\ref{fig_1}b). When observing the red circles shown in Fig.~\ref{fig_6}e2,f2, we can further find that the result from PgNN effectively eliminates the alias effect. 

As a summary, PgNN can perfectly reconstruct high-resolution, wide FOV sample images under general conditions, such as low-defocus condition, and provides a new perspective towards recovered images against traditional methods.

%Further, we apply PbNN to this dataset, and it turns out that original framework performs poorly since the reconstructed color image seems ambiguous and worthless (Fig.~\ref{fig_7}b1). Reason for this phenomenon lies in the mechanisms we incorporated in PgNN. Observing the phase map shown in Fig.~\ref{fig_7}b2, we can find that this image performs well in whole but contains irregular artifacts in the right bottom. Compared to Fig.~\ref{fig_7}c2, on the premise that we provide prior CTF, we can conclude that the TV function promotes to eliminate noise of corrugated shapes shown in Fig.~\ref{fig_7}b2 by making trading off between amplitude and phase.

\begin{figure*}[!t]
	\begin{center}
		\includegraphics[width=0.95\textwidth]{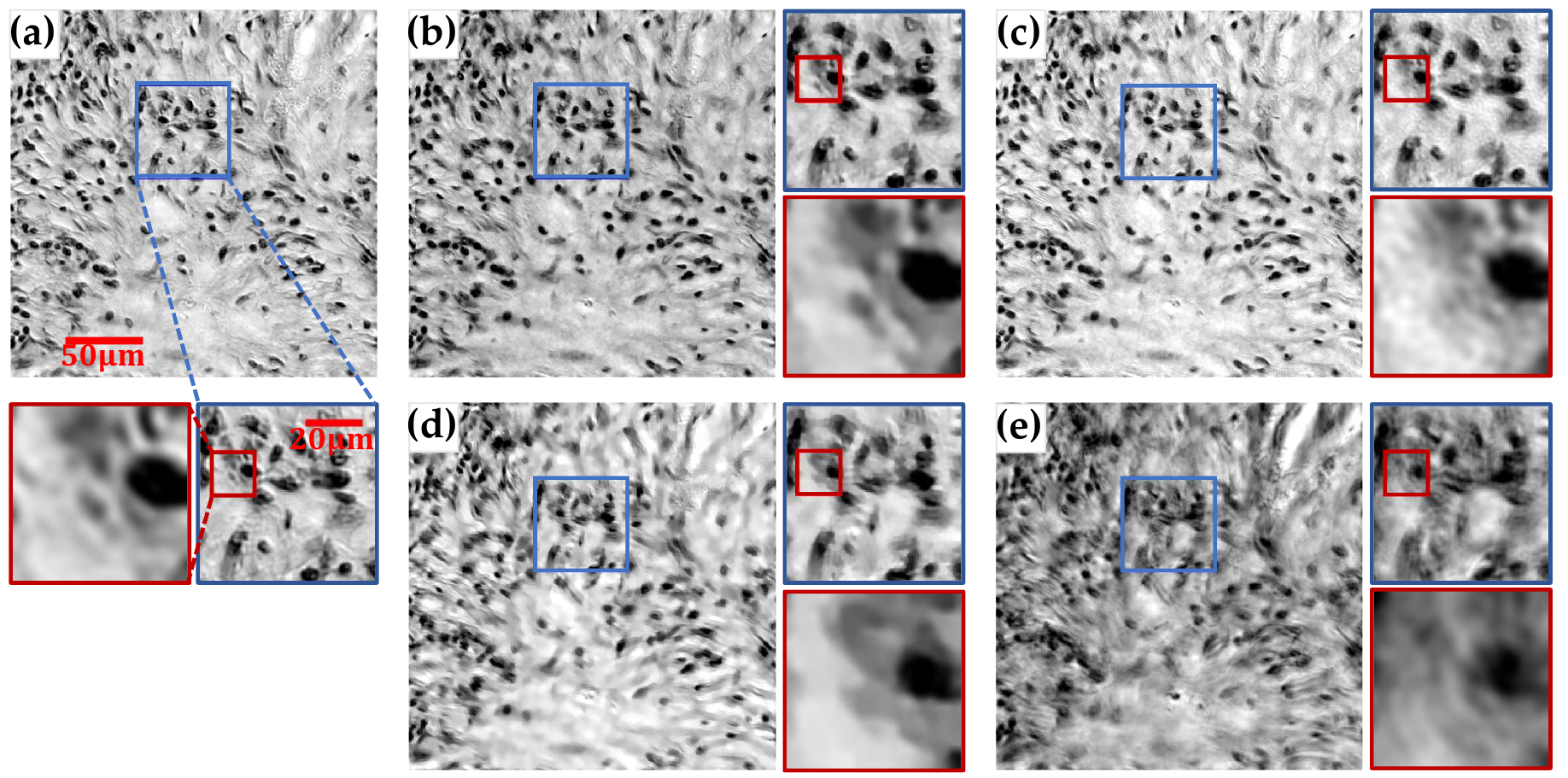}
	\end{center}
	\caption{Recovered intensity images on a tissue slide captured in a high-defocus, high-exposure environment. (a) High-resolution ground truth. (b), (c), (d) Recovered amplitudes from PgNN under different conditions, where (b) uses the complete PgNN, (c) uses an PgNN without TV function and (d) uses an PgNN without Zernike mode. (e) Recovered amplitude from ePIE.}
	\label{fig_8}
\end{figure*}

\begin{figure*}[!t]
	\begin{center}
		\includegraphics[width=0.95\textwidth]{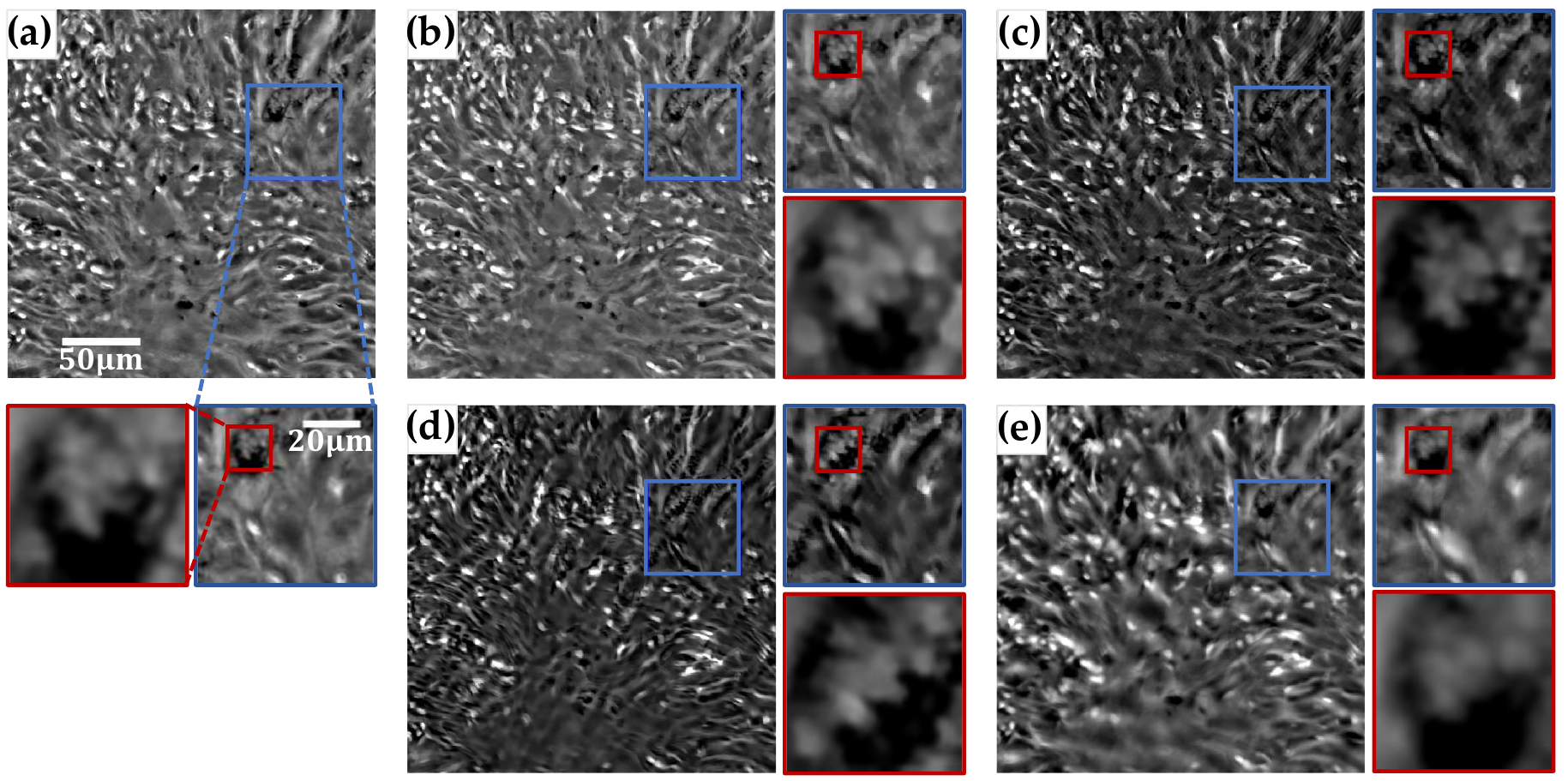}
	\end{center}
	\caption{Recovered phase images in high-defocus, high-exposure environment. The experiment settings are the same as those in Fig. \ref{fig_8}.}
	\label{fig_9}
\end{figure*}

\textbf{Extreme condition.} Although the image reconstruction capabilities of PgNN and ePIE are similar in generic datasets, the Zernike mechanism introduced by PgNN greatly enhances the potential to alleviate the optical aberration. We deliberately collect a tissue section dataset in a high-defocus, high-exposure environment, in which the system aberration seriously affect the qualities of captured intensity images with some pictures being overexposed. To make our results more credible, we additionally test PgNN under different network settings, and adjust the number of Zernike modes $L$ to 50. 

The reconstructed intensity images with some magnified areas are shown in Fig.~\ref{fig_8} (the ground truth is obtained from the corresponding low-defocus dataset). Firstly, compared with the ground truth (Fig.~\ref{fig_8}a), we can easily find that ePIE cannot recover a high-resolution amplitude with distinguishable cell structures (Fig.~\ref{fig_8}e), which also happens in the reconstruction of PgNN without a Zernike function (Fig.~\ref{fig_8}d). Focusing on the center of the secondary magnification area, the PgNN without a TV function (Fig.~\ref{fig_8}c) seems smooth in whole, but is actually rough in detail due to the interference of noise. Meanwhile, the information of small cellular structure is totally lost. In contrast, the complete PgNN can distinguish the small cell characteristics from surrounding tissue structures, and the overall image is smooth with perfect performance (Fig.~\ref{fig_8}b). 

Similar conclusions can be drawn when observing phase components shown in Fig.~\ref{fig_9}. First, images from the PgNN without a Zernike mode (Fig.~\ref{fig_9}d) and ePIE (Fig.~\ref{fig_9}e) are highly blurred and cannot provide instructive information for biomedical applications. Next, from the primary magnification area shown in Fig.~\ref{fig_9}c, it happens that there exist stripe artifacts on the top which can be eliminated by the TV loss (Fig.~\ref{fig_9}b). Finally, by comparing the complete PgNN (Fig.~\ref{fig_9}b) with the ground truth (Fig.~\ref{fig_9}a), we can find that their performances are similar in most cell architectures. 

%In addition, we should note that the amplitude and phase components of the tissue section behaves quite different. We cannot find the corresponding cell structure displayed in the secondary magnification area of phase components (Fig.~\ref{fig_9}) in the intensity images (Fig.~\ref{fig_8}), and this explains the reason why we prefer to recover the complex sample in biomedical algorithms. At the end, we can demonstrate that PgNN is more robust and outperforms ePIE in extreme condition.

%% file: 6.7.discussion_and_summary.tex
\section{Discussion}
The physics-guided model differs from traditional analytical methods and data-driven models in many aspects. Compared to traditional methods, PgNN is much more flexible. By adjusting the learning rate and other hyperparameters, different output results can be obtained. At the same time, the unique backpropagation mechanism of the neural network makes the fusion of principles possible, helping to improve the imaging performance (Fig.~\ref{fig_5}, Fig.~\ref{fig_8}, Fig.~\ref{fig_9}) and make the model more robust (Fig.~\ref{fig_8}, Fig.~\ref{fig_9}). Compared to the data-driven networks, PgNN also has several advantages. First, the proposed physics-guided model enhances the interpretability of the neural network, since it is designed to approximate the forward imaging process of FP. Second, it reduces the required amount of training data (only the low-resolution images captured under the varying illumination angles are required in our experiments). Third, our PgNN can reconstruct the image via unsupervised learning. In summary, by incorporating principles and theories from deep learning and optic imaging, PgNN can be oriented to perform reconstruction by changing the backward dataflow and outperform traditional methods.

It should be noted that one of the motivations of our work is to address the difficulties in obtaining biomedical datasets, which are expected to be resolved in the near future. We used well-designed model and specific principles such as the TV regularization and Zernike mode to temporarily fix this drawback, and we will extend our work with the data-driven method in the future work.

\section{Summary}
In this work, we have provided a physics-guided neural network to solve Fourier ptychographic imaging. By gradually incorporating the total variation mechanism and Zernike polynomial function, PgNN can unroll the object image in the complex domain with high resolution and wide FOV. We verified the validity of our model in simulated and experimental datasets, and demonstrated that PgNN was more robust than traditional FP methods in the high-defocus, high-exposure dataset.

At present, PgNN is mainly composed of several valuable principles taken from optic imaging and deep learning so that it can work well in small datasets. It is valuable to introduce the data-driven method in the physics-guided framework to make it more powerful. For example, we can model part of the Fourier ptychographic forward imaging process via an advanced network architecture, like ResNet, FPN, U-net and so on. Consequently, we can use large data to train the network to learn advanced features at the beginning, and then utilize the remaining physics-guided framework to fine tune in new datasets. Next, the introduction of the TV regularization and Zernike mode helps to improve the network performance. Although we can foresee the effect of different coefficient combinations, manual adjustment is still required for redundant operation in practice. In addition, since there exist no general automatic indicators for biomedical applications, how to optimize the image evaluation system, with which the network can automatically adjust the performance, is also a feasible direction of research.